\documentclass[aps,prx,amsfonts,reprint,superscriptaddress,showpacs,showkeys,nofootinbib]{revtex4-1}

\usepackage{graphicx}
\usepackage{booktabs}
\usepackage{topcapt}
\usepackage{float}
\usepackage{amsmath}
\usepackage{amssymb}
\usepackage{amstext}
\usepackage{mathrsfs}
\usepackage{latexsym}
\usepackage{float}
\usepackage{subfigure}
\usepackage{wasysym}
\usepackage{fancyhdr}

\begin{document}

\title{Experimental observation of $\beta$-delayed neutrons from $^{9}$Li as a way to study short-pulse laser-driven deuteron production}
\author{A. Favalli}
\affiliation{Los Alamos National Laboratory, Los Alamos, New Mexico 87545, USA}
\author{N. Guler}
\affiliation{Los Alamos National Laboratory, Los Alamos, New Mexico 87545, USA}
\affiliation{Spectral Sciences, Burlington, Massachusetts 01803, USA}
\author{D.Henzlova}
\affiliation{Los Alamos National Laboratory, Los Alamos, New Mexico 87545, USA}
\author{K. Falk }
\affiliation{Los Alamos National Laboratory, Los Alamos, New Mexico 87545, USA}
\affiliation{ELI Beamline, Institute of Physics of the ASCR, Na Slovance 2, Prague 182 21, Czech Republic}
\author{S.Croft} 
\affiliation{Oak Ridge National Laboratory, 1 Bethel Valley Road, Oak Ridge, TN 37831, USA}
\author{D.C.Gautier}
\affiliation{Los Alamos National Laboratory, Los Alamos, New Mexico 87545, USA}
\author{K.Ianakiev} 
\affiliation{Los Alamos National Laboratory, Los Alamos, New Mexico 87545, USA}
\author{M.Iliev}
\affiliation{Los Alamos National Laboratory, Los Alamos, New Mexico 87545, USA}
\author{S.Palaniyappan} 
\affiliation{Los Alamos National Laboratory, Los Alamos, New Mexico 87545, USA}
\author{M.Roth}
\affiliation{Institut f\"ur Kernphysik, Technische Universität Darmstadt, Schloßgartenstrasse 9, D-64289 Darmstadt, Germany}
\author{J.Fernandez}
\affiliation{Los Alamos National Laboratory, Los Alamos, New Mexico 87545, USA}
\author{M. Swinhoe}
\affiliation{Los Alamos National Laboratory, Los Alamos, New Mexico 87545, USA}

\begin{abstract}

A short-pulse laser-driven deuteron beam is generated in the relativistic transparency regime and aimed at a beryllium converter to generate neutrons at the Trident laser facility. These prompt neutrons, primarily coming from deuterium break-up in the converter, have been used for active interrogation to detect nuclear materials, the first such demonstration of a laser-driven neutron source. During the experiments, delayed neutron production from $^9$Li decay was observed. It was identified by its characteristic half-life of 178.3 ms. Production is attributed to the nuclear reactions $^9$Be(d,2p)$^9$Li and $^9$Be(n,p)$^9$Li inside the beryllium converter itself. These reactions have energy thresholds of 18.42 and 14.26 MeV respectively, and we estimate the (d,2p) reaction to be the dominant source of $^9$Li production. Therefore, only the higher-energy portion of the deuteron spectrum above the threshold contributes to the production of the delayed neutrons. It was observed that the delayed-neutron yield decreases steeply with increasing distance between the converter and the deuteron source. This behavior is consistent with deuteron production with energy greater than $\sim$20 MeV within a cone with a half-angle greater than 40$^{\circ}$. Measurements on axis with the neutron time-of-flight diagnostic of the prompt neutrons at varying separation between the converter and laser target indicate that the fast deuteron population above threshold is severely depleted on axis out to $\sim$20$^{\circ}$. These measurements are consistent with emission of the fast deuterons (i.e., above 10 MeV/nucleon) in a ring-like fashion around the central axis. Such an inferred ring-like structure is qualitatively consistent with a documented signature of the breakout afterburner (BOA) laser-plasma ion acceleration mechanism. The measurement of $\beta$-delayed neutrons from $^9$Li decay could provide an important new diagnostic tool for the study of the features of the deuteron production mechanism in a non-intrusive way. Experimental measurements, currently not available, of the $^9$Be(d,2p) cross-section are needed to enable quantitative comparison with theoretical models.

\end{abstract}

\maketitle

\section{Introduction} Intense laser-driven ion beams have been the subject of considerable study for over a decade \cite{1,2}. Based on advanced mechanisms of laser-driven ion acceleration,  a new intense and short-duration neutron source with record flux ($> 10^{10}$ n/sr) \cite{3,4} has been pioneered at Los Alamos National Laboratory (LANL). The neutrons are generated from a multistep process starting with the interaction of a short-pulse laser with a deuterated-plastic nanofoil target to make an intense beam of protons and deuterons. The ion beam subsequently impinges on a suitable converter material to drive the neutron beam. This source has the particularly useful properties of high intensity, short-duration and a forward peaked distribution. Laser-driven neutron sources offer an alternative path for the development of compact, bright and penetrating sources for many applications \cite{5}. One of the motivations at LANL for such a source is the capability to perform an assay of special nuclear materials for nuclear materials accountancy, safeguards and national security applications. This source is particularly suitable for the latter two applications. A penetreting intense neutron burst offers the potential of achieving a high signal-to-noise ratio in difficult environments (e.g., with high neutron background emitted by the interrogated item) and promises a short assay time, which translates to a high interrogated item throughput. This application, also known as active interrogation, is based on the measurements of induced neutron signatures to identify/assay nuclear materials during (prompt fission neutron) and after (delayed neutrons from fission products) an interrogation with an external neutron pulse \cite{6,7}, such as the laser driven neutron source at Trident. 

Laser-driven ion acceleration relies on very intense ($I > 10^{18}$ W/cm$^2$) laser fields on plasma targets to create collective effects that drive large accelerating electric fields with field strengths of tens of TV/m over very short distances (microns). The LANL Trident laser facility \cite{8} provides a very high-contrast laser pulse of energy $\sim$80 J, wavelength $\lambda_0 = $1053 nm, ~600 fs FWHM duration, and peak on-target laser intensities up to 10$^{21}$ W/cm$^2$. The high-contrast enables fielding plastic-foil targets ($\sim$1 g.cm$^{-3}$) with thicknesses typically in the range of 300$-$700 nm. The target heats up rapidly during the main pulse rise and becomes relativistically transparent \cite{9} to the laser by the time of peak power. After transparency, the laser interacts volumetrically with the plasma and accelerates ions in the interaction region to high energies \cite{10,11,12,13,14}. 
Recent experiments have demonstrated that high-energy deuterons produced in the relativistically transparent regime can create a forward directed neutron emission primarily from deuteron break-up mechanisms in the converter, in addition to the prompt isotropic component from reactions such as $^9$Be(d,n)$^{10}$B \cite{4}. 

\begin{figure}[htb!]
	\begin{center}
	\includegraphics[width=1\columnwidth]{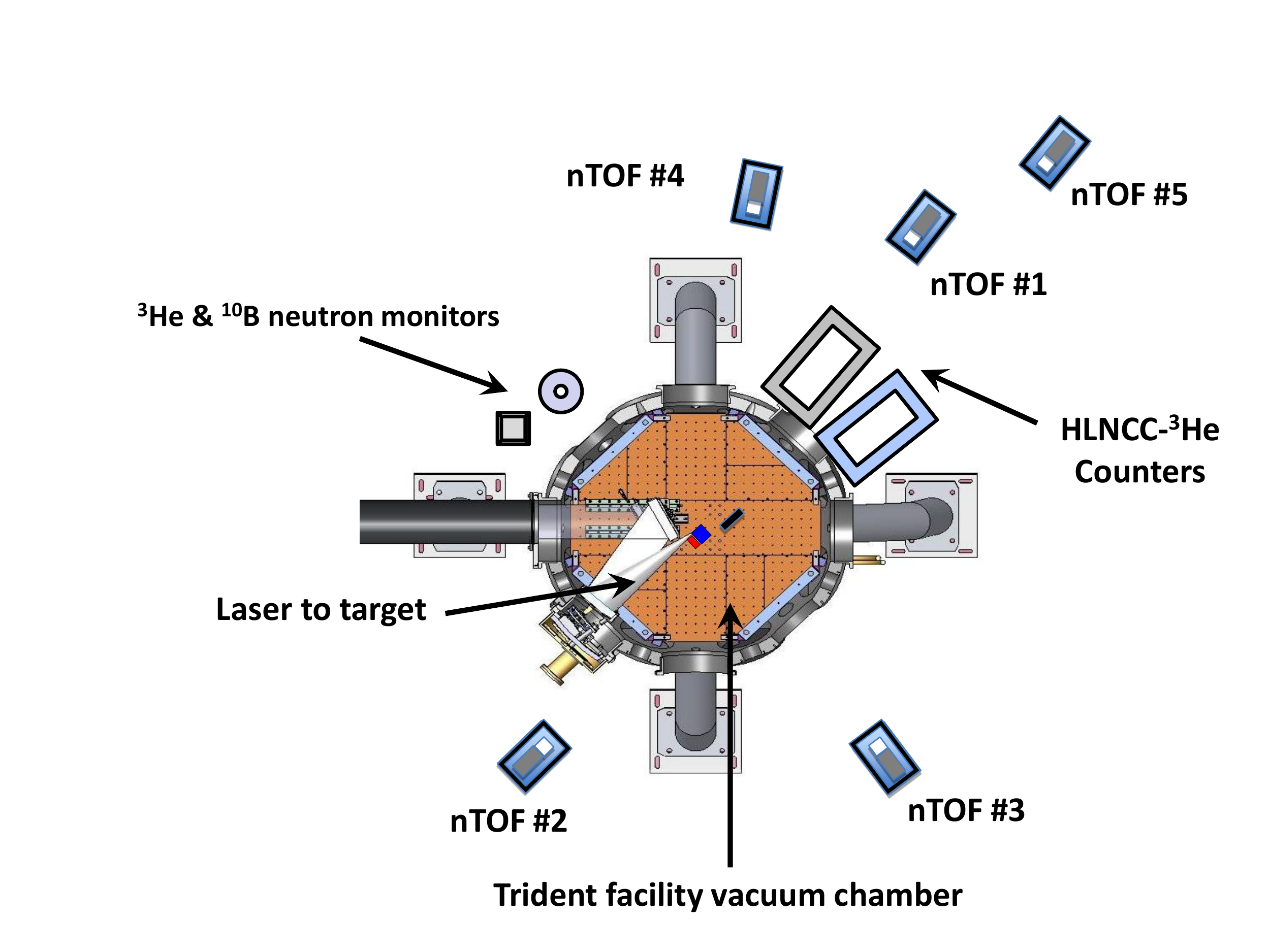}
	\caption{Experimental setup at Trident facility  (not in scale).}
	\label{fig:1}
	\end{center}
\end{figure}

We show for the first time that $\beta$-delayed neutron production from certain nuclear reactions is useful to probe the deuterium beam. We consider two distinct such reactions that result in $^9$Li: $^9$Be(d,2p)$^9$Li \cite{21} and $^9$Be(n,p)$^9$Li \cite{22}. $^9$Li subsequently $\beta$-decays, with one branch producing delayed neutrons with the 1/e-value of $\tau =$ 257.2 millisecond (half-life of 178.3 ms) \cite{15}. As discussed below, only high-energy deuterons are implicated in generating $^9$Li because of the energy thresholds for these reactions. In this letter, we discuss the angular distribution of high-energy deuteron production in the relativistic transparency regime that are revealed through the analysis of the delayed neutron production from $^9$Li decay and of the prompt neutron spectra along the axis. 

\section{Setup} 
Our experimental setup is shown in Fig.~\ref{fig:1}. The target chamber houses the laser focusing optics (in this experiment a F/3 parabolic mirror), the target (for this run, primarily $\sim$350 nm thick deuterated polyethylene (CD$_2$) foils) and the converter (a Be cylinder 20 mm diameter and 40 mm depth). All the diagnostic equipment, described below, was positioned outside the chamber. The chamber radius is 1 m and its wall is 20 mm thick stainless steel, with 25 mm thick Al flanges located around the chamber. The neutron diagnostic arrangment was motivated to utilize laser-driven neutron beams for active interrogation. Two $^3$He thermal neutron coincidence well counters were placed outside the chamber for the detection of $\beta$-delayed neutrons from active interrogation of nuclear material. These arise from neutron rich nuclei created following induced fission. One of the well counters contained a sample of nuclear material for active interrogation, while the other was kept empty to serve as a reference for background comparison. The counters used were high-level neutron coincidence counters (HLNCC-II) \cite{16} composed of a single ring of 18 $^3$He-filled proportional detectors embedded in high-density polyethylene. Neutrons impingings on the HLNCC-II detectors are counted in pulse mode being spread out in time by the thermalization and diffusion property of the moderator assembly. Both counters were located on the equator, closely straddling the central beam axis, defined by the laser propagation direction. It is important to emphasize that for the purpose of this letter, only data acquired in the reference (empty) detector are considered for the analysis, where no fissionable material that could lead to delayed neutron production is present. In addition, a single moderated $^3$He-filled proportional detector, was located $\sim$90$^{\circ}$ off-axis. Furthermore, 5 neutron time-of-flight (nTOF) plastic scintillator detectors were positioned around the target chamber. One nTOF detector (nTOF \#5) was located 6.2 m away from the converter along the central beam axis (the laser-propagation direction and the symmetry axis of the converter). This detector was used to measure the neutron energy distribution in the forward direction. It was located outside the building to detect the high-energy portion of the neutron spectrum with better resolution. Several bubble detectors were distributed around the chamber to measure laser generated neutron flux in multiple directions \cite{3,4,5}.

The moderated $^3$He detector, as well as the HLNCC-II well counter, were covered in a thin ($\sim$1 mm thick) Cd foil to block the contribution of slow neutrons returning to the detectors after scattering in the room \cite{17}. Prompt beam-neutrons produced by the laser that strike these detectors exhibit a characteristic 1/e die-away time of several tens of microseconds due to the thermalization and diffusion in the high-density polyethylene moderator \cite{17}. The die-away times of HLNCC-II and the single moderated $^3$He detector correspond to $\sim$43 and $\sim$20 $\mu$s, respectively. These time intervals are short compared to the delayed neutron production which typically extends over several milliseconds or more. Thus, the delayed neutron signal can be clearly identified in these thermal-neutron detection systems. The single $^3$He detector positioned off the beam-axis was used as a neutron flux monitor, since its die-away time characteristics enabled data collection over an extended period of time beyond the initial time window, when the well counter was recovering from the initial neutron burst. The count rates from both detectors were acquired using a list mode (time stamp)card with 10 ns time resolution and analysed in the form of time-interval distributions of the neutron detection times following the laser trigger pulse.

\section{Experiment} 
In the initial phase of these experiments, various measurements were made to empirically optimize the neutron production and its directionality for active interrogation. The characteristics of the neutron beam were investigated by changing target/converter configurations and varying the distance between them in a series of laser shots. The distance of the front face of the Be converter to the target was systematically increased to 3.6, 6.0, 8.0 and 12.0 mm over several shots and the results are reported here. 

\begin{figure}[htb!]
	\begin{center}
	\includegraphics[trim=0cm 0cm 1cm 0cm,clip=true, width=0.8\columnwidth]{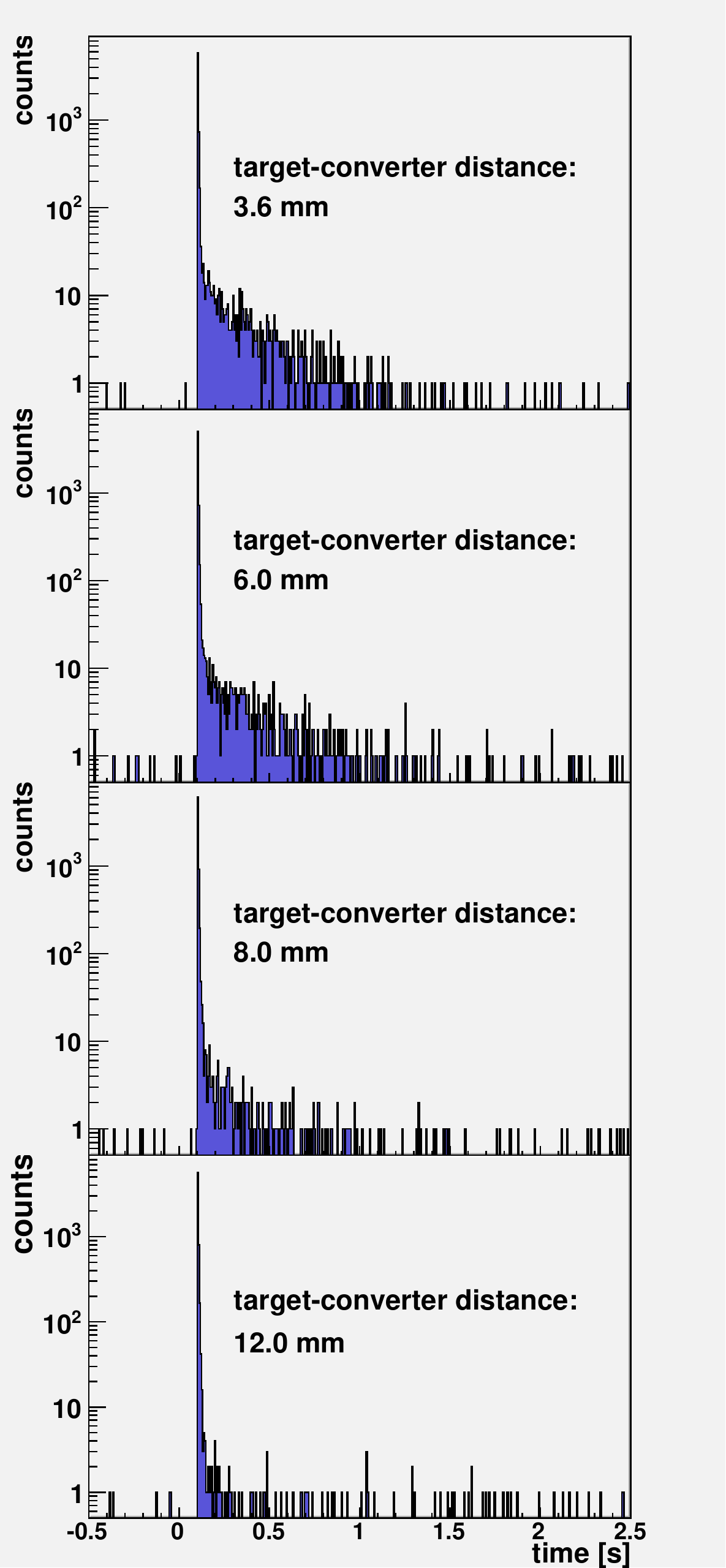}
	\caption{Delayed neutron production in the forward direction from $^9$Li decay is shown for four different shots in which Be ion-to-neutron converter was placed at different distances from the target as measured with the HLNCC-II detector. As the converter gets closer to the target, more high energy deuterons interact with the converter, hence increasing the $^9$Li production, and the subsequent delayed neutron decaying signal. These are the raw data before normalization for neutron yield. Note that zero on the x-axis corresponds to the time of laser pulse and negative values represent background data acquired immediately before the shot.}
	\label{fig:2}
	\end{center}
\end{figure}

\begin{figure}[htb!]
	\begin{center}
	\includegraphics[trim=0cm 0cm 1cm 0cm,clip=true, width=0.8\columnwidth]{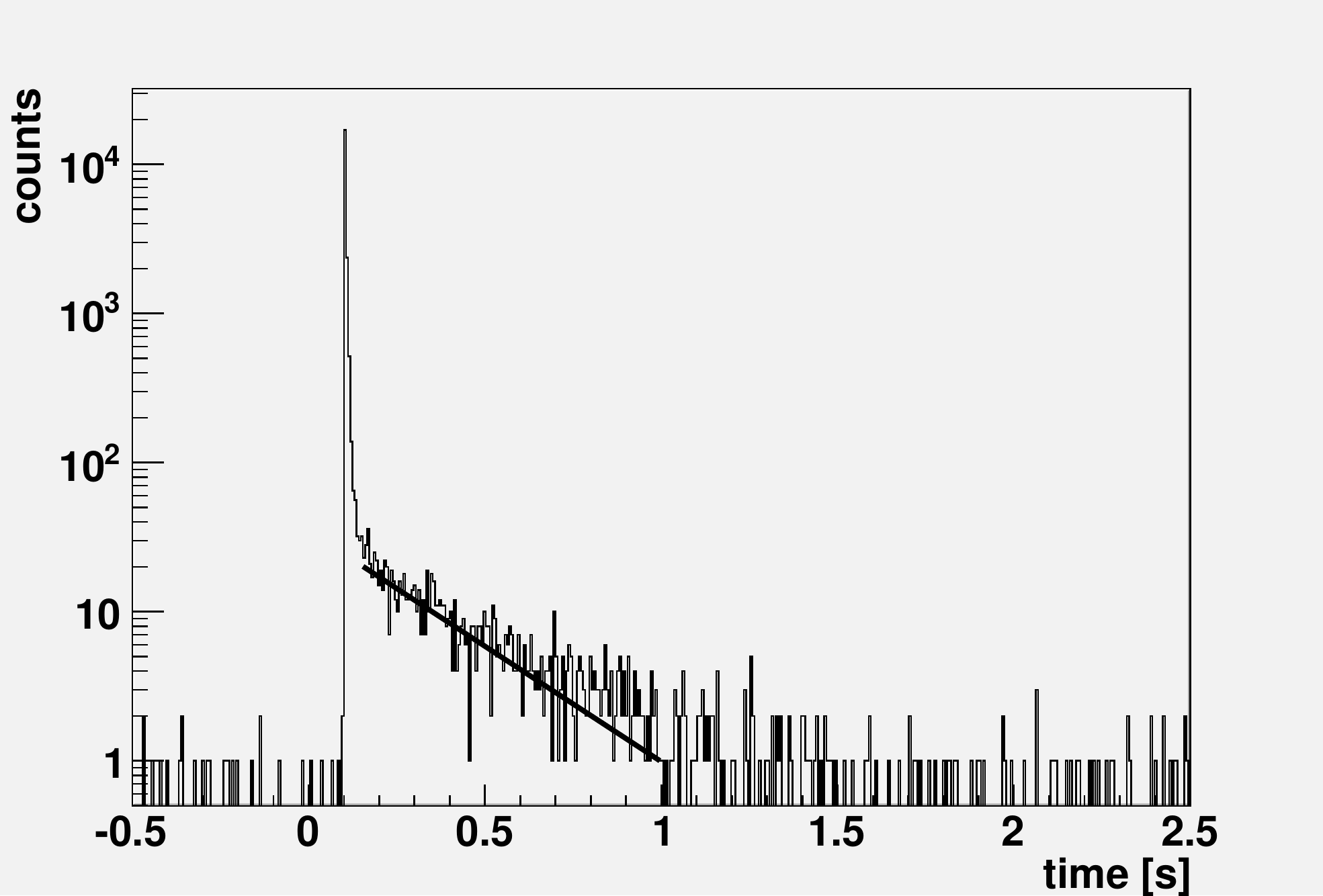}
	\caption{Summed time-interval distributions measured in the HLNCC counter for target/converter distances of 3.6, 6.0 and 8.0 mm. The decay constant (260 $\pm$ 20 ms ) is calculated by a weighted least squares fit and the uncertainty (1$\sigma$) by varying the fit region.}
	\label{fig:3}
	\end{center}
\end{figure}

\begin{figure}[htb!]
	\begin{center}
	\includegraphics[trim=0cm 12cm 1cm 0cm,clip=true, width=0.8\columnwidth]{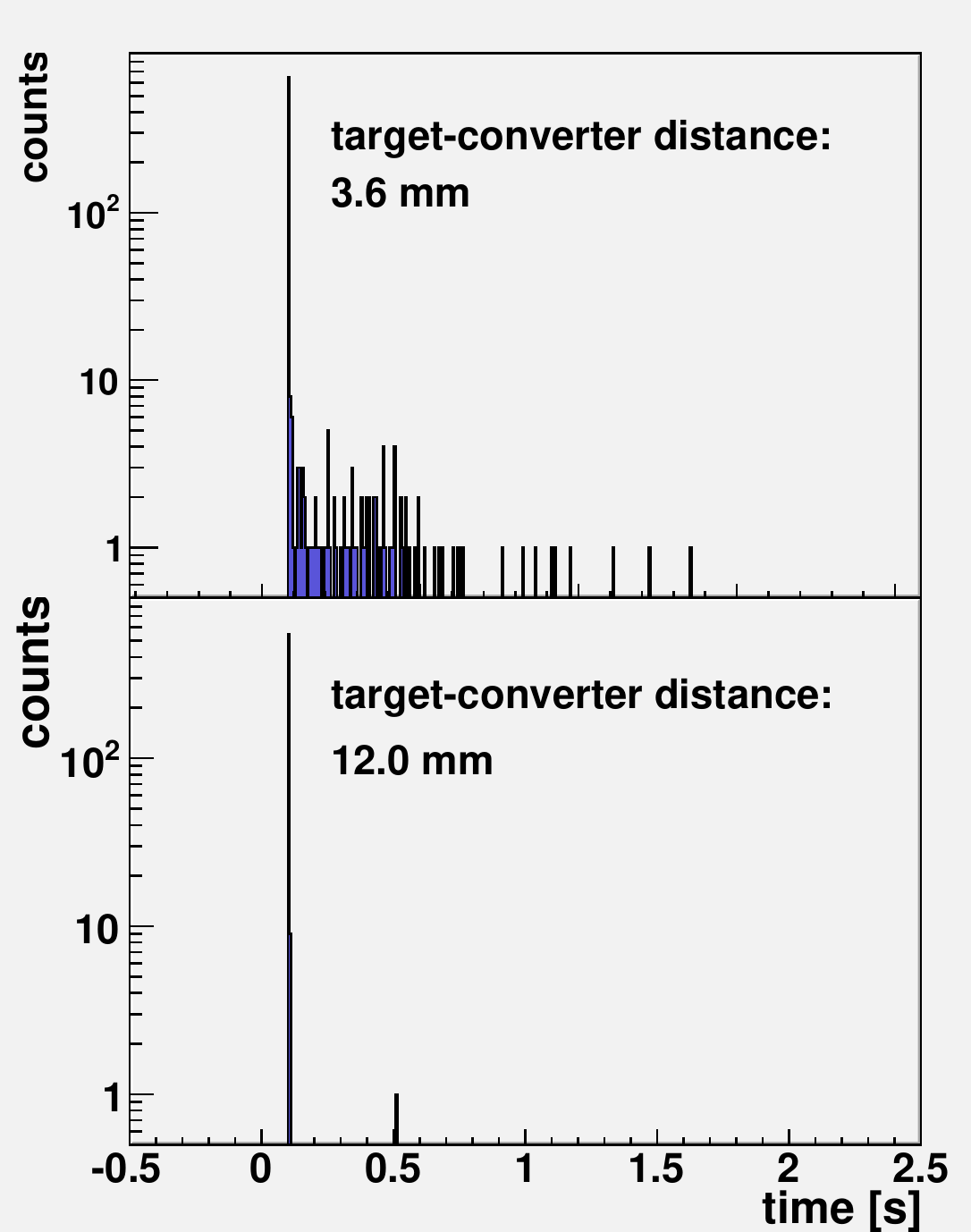}
	\caption{$^9$Li delayed neutron spectra in the $^3$He neutron monitor located at $\sim$90$^{\circ}$ from the forward direction shown for two shots with different converter-target distances. The $^9$Be converter catches high energy deuterons when at 3.6 mm, while at 12 mm, it misses most them. Delayed neutron from $^9$Li decay is expected isotropic, thus consistently the $^9$Li delayed neutron signal is measured by detectors located at $\sim$90$^{\circ}$.}
	\label{fig:4}
	\end{center}
\end{figure}

During the measurements in which the Be converter was placed at a separation of either 3.6, 6.0 or 8.0 mm, an unexpected delayed-neutron tail following the prompt neutron peak was observed in the $^3$He detectors not containing any nuclear material. So the origin of these neutrons cannot be from induced fission. The tail was observed in the reference HLNCC-II well counter located in the forward direction as well as in the single $^3$He detector located at $\sim$90$^{\circ}$ off-axis, as shown in Figs.~\ref{fig:2} and \ref{fig:4}, respectively. These Figures show the time-interval distribution of the neutron detection recorded following the laser pulse. It can be seen that the tail decays completely in $\sim$1 second. The 1/e-value ($\tau$) extracted from these experiments  corresponds to 260 $\pm$ 20 ms. This value was obtained by summing-up the  time-interval distributions measured in  the HLNCC-II well counter for target-to-converter distances of 3.6, 6.0 and 8.0 mm and performing an exponential fit to determine the slope of the decay (see Fig.~\ref{fig:3}). The extracted 1/e-value ($\tau$) corresponds to the delayed neutron production via decay of $^9$Li with the 1/e-value of $\tau =$ 257.2 ms \cite{15}. This delayed neutron production is expected to be isotropic, as confirmed by the comparison of the signals from the HLNCC-II and the $^3$He flux monitor, located at $\sim$0$^{\circ}$ and $\sim$90$^{\circ}$, respectively.

Because of the die-away time characteristics of the $^3$He detectors used, the temporal signal produced directly by the laser-driven prompt neutron pulse dissipates approximately within a ms, since 1 ms corresponds to more than 20 times the 1/e-die-away period. In addition, return of the scattered thermal neutrons from the room is suppressed by the external Cd shield. Contribution of these effects to the delayed neutron signature, observed over the period of 50$-$1000 ms, can therefore be excluded. 

The energy thresholds of the nuclear reactions $^9$Be(d,2p)$^9$Li and  $^9$Be(n,p)$^9$Li are 18.42 and 14.26 MeV, respectively \cite{17}. Therefore, the delayed neutrons can only be produced from the high-energy deuterons or neutrons impinging on the $^9$Be converter. The high-energy deuterons with energies $\ge$ 18.42 MeV can lead to $^9$Li production directly via $^9$Be(d,2p)$^9$Li. In addition, the deuterons with energies greater than about 30 MeV can also contribute, by break-up reactions, to the production of the high-energy neutrons, that lead to $^9$Li production by the (n,p) reaction. The conversion efficiency of deuterons-to-neutrons for the deuterons above 15 MeV has been measured to be around 0.1\% \cite{4} for the same size Be converter. The low conversion efficiency combined makes the $^9$Be(d,2p)$^9$Li branch the dominant process.

\section{Results} 
\begin{figure}[htb!]
	\begin{center}
	\includegraphics[width=1\columnwidth]{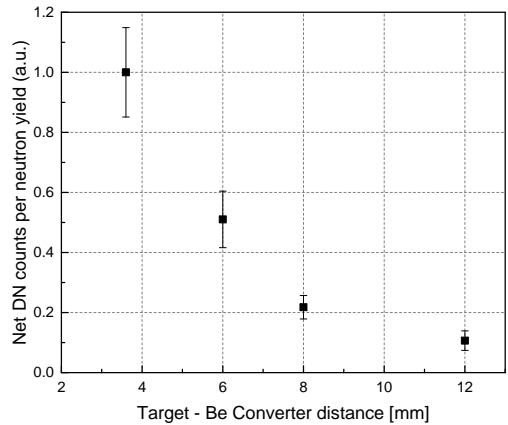}
	\caption{$^9$Li delayed neutron (DN) production, normalized for the neutron yield (HLNCC-II at 0$^{\circ}$), as a function of the distance between the Be back face and the target. The uncertainties are statistical only. The uncertainty on the 3.6 mm datum is larger then the others because it contains a correction factor for the orientation axis of the HLNCC-II detector.}
	\label{fig:6}
	\end{center}
\end{figure}

We observed that as the distance of the converter to the CD$_2$-target increases, the yield of the delayed neutrons, proportional to $^9$Li production, decreases. 
Figs.~\ref{fig:2} and \ref{fig:4} illustrates this trend. Therefore, at a large enough target-to-converter distance, the flux of the high-energy deuterons on the converter, that govern the $^9$Li production, approaches zero. 
The integral of the delayed neutron counts normalized to the laser neutron yield are shown in Fig. \ref{fig:6} for all the target-converter distances. The delayed neutron counts displayed were integrated over an interval of 50$-$1000 ms after the laser pulse. It can be anticipated that at a sufficiently close distance between the converter and the deuteron source most of the deuterons would hit the Be and the dependence versus separation would level off. Such distances (less than 3.6 mm) were not investigated in this experiment in order to avoid Be damage and contamination inside the chamber. These observations and their scaling imply a large divergence angle for the high-energy deuteron production responsible for the $^9$Li (see Discussion). Although these data enable us to estimate a bound for the cone angle of the fast deuterons, they don't provide information about the energy distribution within that cone, except that it exceeds the $\sim$20 MeV threshold.

\begin{figure}[htb!]
	\begin{center}
	\includegraphics[width=0.9\columnwidth]{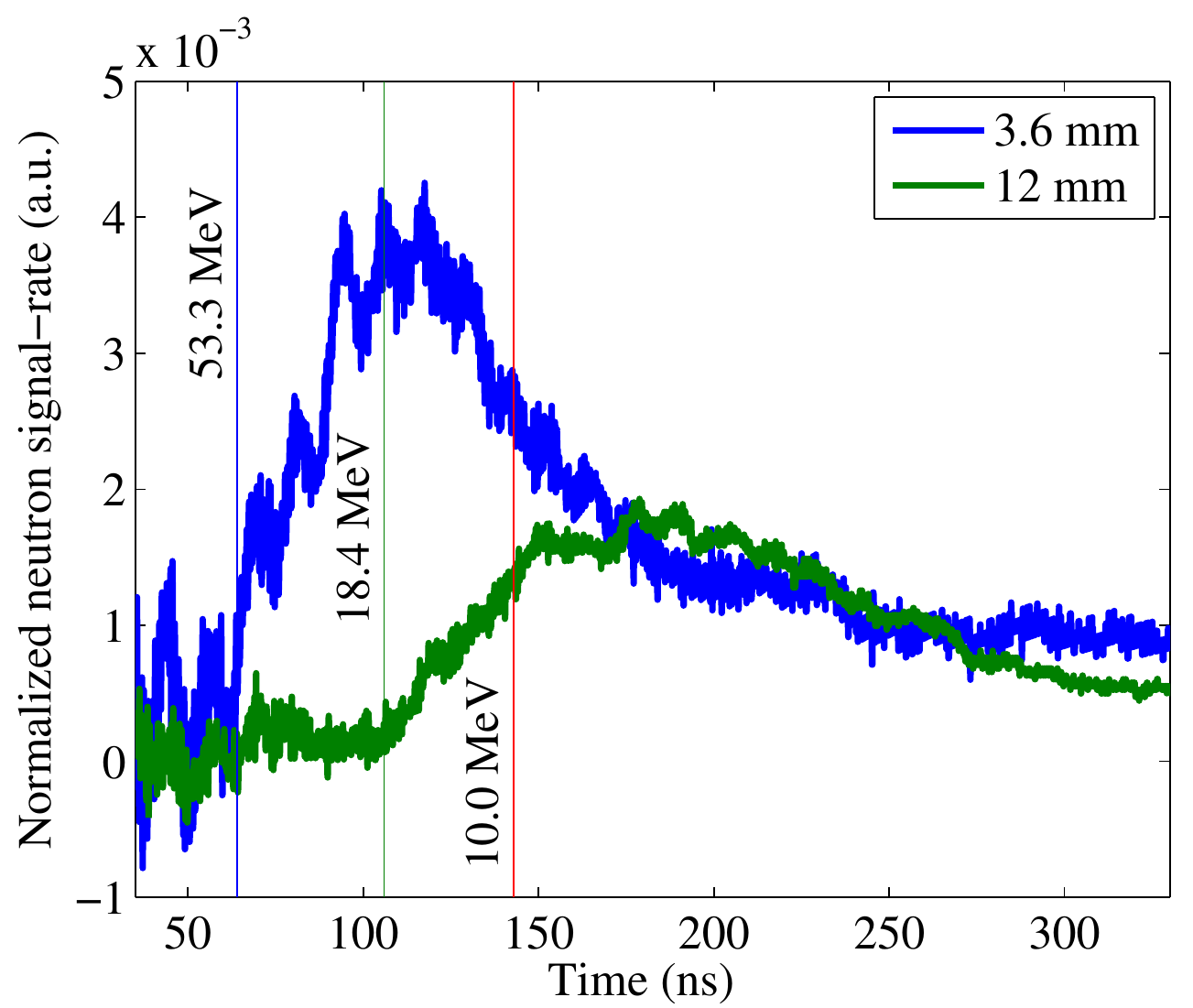}
	\caption{Neutron time-of-flight spectra (nTOF \#5) for two shots with the Be converter placed at 3.6 and 12.0 mm from the target, respectively. The vertical lines mark the highest observed energy of the neutron distribution of $\sim$55 MeV and $\sim$18 MeV for 3.6 mm and 12 mm distances, respectively.  A further vertical line marks 10 MeV neutron energy. Higher energy neutrons (to the left of this line) can only be produced by breakup of high energy ($>$20 MeV) deuterons.}
	\label{fig:5}
	\end{center}
\end{figure}

Independent data from the nTOF detector located on the central beam axis (nTOF \#5) provides insight on the energy distribution of the fast deuterons within the cone discussed above. nTOF \#5 data in Fig.~\ref{fig:5} show the dependence of the neutron beam energy on the target-to-converter distance. Specifically, it shows time-of-flight spectra  of the prompt neutron beam arising from the deuterium disintegration in the converter for target-to-converter distances of 3.6 and 12 mm. The deuterons of interest ($>$ 20 MeV), when undergoing break-up, produce neutrons with energies of $\sim$10 MeV or more. Therefore in Fig.~\ref{fig:5}, we compare the neutron spectra for times 143 ns (corresponding to 10 MeV) and lower (corresponding to higher energies). If the deuteron spectra were the same at all angles we would expect the prompt neutron spectra in nTOF\#5 to be self-similar for any separation. Instead, Fig.~\ref{fig:5} shows a spectrum severely depleted of neutrons above 10 MeV when the converter separation is 12 mm, measured relative to the spectrum at the minimum separation of 3.6 mm. This suggests an angular distribution of the fast deuterons that is hollow along the central axis (see discussion below.)

\section{Discussion} 
If the observed delayed neutron production from $^9$Li decay was caused by a mechanism involving a quasi-parallel beam of forward directed deuterons above the $^9$Li production threshold striking the Be converter, the delayed neutron yield would be comparatively insensitive to the distance between the CD$_2$ laser target and the Be converter. This is inconsistent with our observations. Instead, our observations and interpretations imply a large divergence angle for the high-energy deuterons responsible for the $^9$Li production. There must be a large number of high energy deuterons that impinge on the Be converter when it is close to the target but miss the converter completely when it is farther away. The minimum cone half-angle relative to the central axis (the axis of the converter cylinder as well as the laser propagation direction) of these fast deuterons can be estimated in two ways. First, we simply take the $\arctan$ of the angle defined by the ratio of the converter radius (1 cm), and the closest separation (0.4 cm), which yields 68$^{\circ}$. To refine this estimate, we note that a representative energetic deuteron ($\sim$50 MeV) \cite{3,4} would penetrate $\sim$0.85 cm into the Be converter. So we could take instead the closest separation $+$0.85 cm, which yields an angle of $\sim$40$^{\circ}$.

As mentioned above, if the spectra of the deuterons driving the prompt neutrons in Fig.~\ref{fig:5} were the same at any angle, we would expect the prompt neutron spectra to be self-similar for any separation between the laser target and neutron converter. Instead, Fig.~\ref{fig:5} shows that the fast deuteron angular distribution cannot be uniform. The lack of neutrons with energies greater than 10 MeV at the larger target-converter distance implies that most high energy deuterons are missing the converter. Using a similar estimation as above, the fast deuteron angular distribution is depleted along the central axis up to an angle $\sim$20$^{\circ}$.  All these observations are consistent with the bulk of the fast deuterons coming out in a ring-like fashion.

The observation of a ring-like morphology for the fast deuteron population is consistent in light of prior simulations and experiments on laser-driven C$^{6+}$ beams in the BOA regime. We note that deuterons have the same charge to mass ratio of fully ionized C. 3D simulations of the interaction of an intense laser with diamond nanofoils resulting in BOA acceleration discussed in Ref. \cite{19} show such a ring-like structure for the high-energy portion of the C$^{6+}$ spectrum. The experiments in Ref. \cite{19} carried out on the Trident laser with diamond nanofoils also showed an increasingly ring-like emission versus angle for increased C$^{6+}$ ion energy, albeit at a smaller angle $\sim$10$^{\circ}$. Although, deuterated plastic and diamond targets are different in their chemical structure, with proper optimization of the target thickness to yield a similar time during the laser pulse when the laser target becomes relativistically transparent, it is not unreasonable to expect similar behavior. However, the response of the two materials to the laser pre-pulse and the laser hydrodynamic disassembly would not be identical, so quantitative differences are expected.

Our results demonstrate the power of selected nuclear reactions as a highly specific diagnostic for deuteron beam spectra and morphology. In the case of Be converter $\beta$-delayed neutron measurements from $^9$Li decay offer an attractive alternative relative to other diagnostics tools, such as radio-chromic films, because they can be measured continuously throughout any experimental campaign, and the converter obstructs and thus complicates the use of other beam diagnostics. As demonstrated in Fig.~\ref{fig:4}, a simple moderated $^3$He-filled proportional counter may be used to detect this delayed neutron signature while providing a flux monitoring capability. The Be converter does not need to be segmented, as the use of radio-chromic films would require, and the data are acquired automatically as part of regular flux monitoring. Nuclear diagnostic methods such as this one, could also be very useful to resolve ambiguous results with particle spectrometers when ions with similar charge to mass ratios are involved. 

The measured integral delayed neutron counts together with known efficiency of the $^3$He detector used and the $^9$Li production cross section, could be used to determine the fraction of high energy deuterons in every shot. In general, the quantitative power of the technique would be greatly increased if the reaction cross section was known as a function of energy. Unfortunately, there is only a theoretical calculation available \cite{20}. To the extent of our knowledge, the $^9$Li production from $^9$Be(d,2p)$^9$Li reaction has not been reported in a peer reviewed publication since 1951 \cite{21}. Therefore, our work provides an impetus for further work to measure the cross section experimentally. In fact, a modified version of the experimental set-up utilized in the current work could be used for cross section measurements. In addition, the observation of $\beta$-delayed neutrons from sources other than nuclear fission has important implications for the correct understanding and interpretation of laser-driven active interrogation measurements, when $\beta$-delayed neutrons from fissions are being used as the signature of nuclear material. In this case, delayed neutron sources other than nuclear fission have to be understood and accounted for to prevent biases in the estimates of the amounts of nuclear material \cite{5}.

\section{Conclusions}
An intense and energetic deuterium beam is generated by irradiating a sub-micron thick polyethylene target with the intense Trident short-pulse laser. The ion acceleration mechanism has been identified as BOA, which operates in the relativistic transparency regime of laser-plasmas. The deuterons produce neutrons by impinging on a beryllium cylinder placed in the beam path. This technique produces a higher forward-directed neutrons flux per unit laser energy than any alternative available so far. Our neutron detectors, which were designed for active interrogation of special nuclear material, unexpectedly detected delayed neutrons from $^9$Li decay with its characteristic half-life of 178.3 ms. We attribute the $^9$Li production primarily to $^9$Be(d,2p) reactions driven by energetic deuterons above the reaction threshold of 18.42 MeV. 
The delayed neutron yield gives a direct on-line measure of the number of high energy deuterons impinging on the beryllium collector. The measured delayed neutron yield from the $^9$Li decay is observed to decrease steeply as the separation between the laser target and neutron converter is increased. That dependence implies that the high energy portion of the deuteron beam is emitted from the laser target in a cone with a half angle (relative to the central axis) of at least $\sim$ 40$^{\circ} - 70^{\circ}$. Moreover, data from the neutron time of flight diagnostic along the central axis have been used to measure the energy spectrum of the prompt beam neutrons created by deuterium disintegration in the converter, for various laser-target to converter separations. Those neutron spectra have been used to estimate the relative abundance of the fast deuterons, i.e., above the $^9$Li production threshold. It is found that the flux of those fast deuterons is severely depleted along the axis, up to an angle $\sim$20$^{\circ}$. These observations are consistent with the hypothesis that the fast deuteron population is being emitted in a ring-like fashion around the central axis. This hypothesis is qualitatively consistent with one of the documented unique signatures of the BOA mechanism. Observation of the delayed neutron production from $^9$Li decay can be a useful diagnostic for neutron production experiments using deuterons on beryllium converters. The utility of this nuclear diagnostic to determine the beam morphology motivates measurements of the $^9$Be(d,2p) cross-section to compare with current model calculations.  Indeed, it is possible to use a modified version of our experimental setup to measure those cross sections. In addition, as outlined above, this unanticipated source of delayed neutrons must be considered in the interpretation of laser-driven active neutron interrogation measurements.
\section{Acknowledgment}
Research reported in this publication was supported by the LANL Laboratory Directed Research and Development (LDRD) program at Los Alamos National Laboratory. The Trident Laser Facility is supported by the NNSA Science and ICF campaigns.

\bibliography{demag}{}

\end{document}